\documentclass[12pt]{article}

\usepackage{graphics}
\usepackage{graphicx}
\usepackage{amssymb}
\usepackage{amsmath}
\usepackage{amsfonts}
\usepackage{cite}

\begin{document}

\begin{titlepage}
\begin{center}

{\Large \bf{Electronic stress tensor analysis of hydrogenated palladium clusters}
}

\vskip .45in

{\large
Kazuhide Ichikawa$^1$, Ayumu Wagatsuma$^1$, Pawe{\l}  Szarek$^2$, \\
Chenggang Zhou$^{3}$, Hansong Cheng$^{3, 4}$ and Akitomo Tachibana$^{*1}$}

\vskip .45in

{\em
$^1$Department of Micro Engineering, Kyoto University, Kyoto 606-8501, Japan\\
$^2$Wroc{\l}aw University of Technology, Institute of Physical and Theoretical Chemistry, \\
Wybrze\.ze Wyspia\'nskiego 27, 50-370 Wroc{\l}aw, Poland\\
$^3$Sustainable Energy Laboratory, China University of Geosciences Wuhan, \\ 
Wuhan 430074, China P.R. \\
$^4$Department of Chemistry, National University of Singapore, Singapore \\
}

\vskip .45in
{\tt E-mail: akitomo@scl.kyoto-u.ac.jp}

\end{center}

\vskip .4in

\date{\today}

\begin{abstract}
We study the chemical bonds of small palladium clusters Pd$_n$ ($n=2$--9) saturated by hydrogen atoms using electronic stress tensor.
Our calculation includes bond orders which are recently proposed based on the stress tensor. 
It is shown that our bond orders can classify the different types of chemical bonds in those clusters.
In particular, we discuss Pd--H bonds associated with the H atoms with high coordination numbers and the difference of H--H bonds in the different Pd clusters
from viewpoint of the electronic stress tensor. 
The notion of ``pseudo-spindle structure" is proposed as the region between two atoms where the largest eigenvalue of the electronic stress tensor is negative and 
corresponding eigenvectors forming a pattern which connects them.
\end{abstract}

{Wave function analysis; Theory of chemical bond; Stress tensor; hydrogenated Pd clusters}

\end{titlepage}

\setcounter{page}{1}


\section{Introduction} \label{sec:intro}

Studying structures and energetics of small palladium clusters is of great importance as the first step toward understanding their catalytic properties.
In Ref.~\cite{Luo2007}, the structures and physical properties of small palladium clusters Pd$_n$  ($n=2$--15) and several larger clusters have been 
studied  using density functional theory (DFT) calculation. 
They have investigated their isomeric structures extensively and found many energetically nearly degenerate isomers. 
In Ref.~\cite{Zhou2008}, based on the lowest energy structures of Pd$_n$ ($n=2$--9) found in Ref.~\cite{Luo2007}, the role of small palladium clusters
in catalyzing dissociative chemisorption of molecular hydrogen has been studied by DFT calculation. 
The results include the structures of the Pd clusters under full hydrogen saturation. 
As for Pd$_6$ cluster, they have reported detailed analysis of sequential H$_2$ dissociative chemisorption starting from bare Pd$_6$ cluster to Pd$_6$H$_{14}$ cluster. 
Their conclusion of this work is that the capacity of small Pd clusters to adsorb H atoms is substantially smaller on average than that of Pt clusters, indicating 
 that Pd nanoparticles are less efficient than Pt nanoparticles in catalyzing dissociative chemisorption of H$_2$ molecules.
Although this may be not so industrially encouraging result for the Pd clusters, the obtained structures have interesting features from the viewpoint of chemical bonds. 
Our paper is a follow-up study of these papers to learn more about nature of chemical bonds in Pd$_n$ ($n=2$--9) and their hydrogen-saturated versions 
using the electronic structures obtained by quantum chemical calculation.

In our analysis of chemical bonds, we use the electronic stress tensor. This method is based on the Regional Density Functional Theory (RDFT) and Rigged Quantum Electrodynamics (RQED)
\cite{Tachibana1999,Tachibana2001,Tachibana2002a,Tachibana2002b,Tachibana2003,Tachibana2004,Tachibana2005,Tachibana2010}
and has been applied to several molecular systems \cite{Szarek2007,Szarek2008,Szarek2009a,Ichikawa2009a,Ichikawa2009b,Ichikawa2010a,Ichikawa2010b}.
Our method includes recently proposed bond orders \cite{Szarek2007} which are defined using the electronic stress tensor. 
One of our purposes is to show their usefulness in the Pd clusters. As far as metallic clusters are concerned, our analysis had been only applied to Pt clusters \cite{Szarek2009a} and Al$_4$ cluster \cite{Ichikawa2010b} so the present analysis can be useful basis for further research using our stress tensor based analysis. 
Special interest in these Pd clusters is that there seems to be H-H bonds within the clusters. Some of these H-H bonds are considered to form after hydrogen molecules are dissociatively adsorbed to the Pd clusters \cite{Zhou2008}.
It would be intriguing to investigate whether these H atoms are bonded from the viewpoint of electronic stress tensor and, if bonded, 
how the bonding nature differs from that of the free H$_2$ molecule.

This paper is organized as follows. In the next section, we briefly explain our quantum chemical computation method. We also describe our analysis method based on the RDFT  and the RQED, including the definition of our bond orders. In Sec.~\ref{sec:results}, we discuss our results. In Sec.~\ref{sec:bo}, we analyze the chemical bonds of the hydrogenated Pd clusters using our bond orders. In Sec.~\ref{sec:stress}, we discuss the chemical bond using the stress tensor with special emphasis on the Pd--H bonds associated with H atom with high coordination number and the bonds between H atoms. 
In Sec.~\ref{sec:beS} we discuss a way to improve our bond order definition by integrating energy density over some area.
We summarize our paper in Sec.~\ref{sec:summary}.

\section{Theory and calculation methods} \label{sec:calc}
\subsection{Ab initio electronic structure calculation}
We perform ab initio quantum chemical calculation for Pd clusters and their hydrides. 
In this work, calculations are performed by {\sc{Gaussian03}} program package \cite{Gaussian03} using density functional theory (DFT) 
with Perdew-Wang 1991 exchange and correlation functional (PW91) \cite{Perdew1992}. 
The 6-31G** basis set with polarization functions \cite{Hehre1972, Gordon1980, Delley1990} has been used for hydrogen atoms and
LanL2DZ effective core potential \cite{Hay1985} for Pd atoms. 
Optimization was performed without imposing symmetry.

\subsection{RDFT analysis} \label{sec:RDFT}
In the following section, we use quantities derived from the electronic stress tensor to analyze chemical bonds of bare and hydrogenated Pd clusters. This method based on RDFT and RQED\cite{Tachibana1999,Tachibana2001,Tachibana2002a,Tachibana2002b,Tachibana2003,Tachibana2004,Tachibana2005,Tachibana2010} provides useful quantities to investigate chemical bonding such as new definition of bond order \cite{Szarek2007,Szarek2008,Szarek2009a}. We briefly describe them below.
(For other studies of quantum systems with the stress tensor in a slightly different context, see Refs.~\cite{Bader1980,Nielsen1983,Nielsen1985,Folland1986a,Folland1986b,Godfrey1988,Filippetti2000,Pendas2002,Rogers2002,Morante2006,Tao2008,Ayers2009}.
See also Refs.~\cite{Ayers2002,Anderson2010} for related discussion on energy density.)

The basic quantity in this analysis is the electronic stress tensor density $\overleftrightarrow{\tau}^{S}(\vec{r})$ whose components are given by
\begin{eqnarray} 
\tau^{Skl}(\vec{r}) &=& \frac{\hbar^2}{4m}\sum_i \nu_i
\Bigg[\psi^*_i(\vec{r})\frac{\partial^2\psi_i(\vec{r})}{\partial x^k \partial x^l}-\frac{\partial\psi^*_i(\vec{r})}{\partial x^k} \frac{\partial\psi_i(\vec{r})}{\partial x^l} \nonumber\\
& &+\frac{\partial^2 \psi^*_i(\vec{r})}{\partial x^k \partial x^l}\psi_i(\vec{r}) -\frac{\partial \psi^*_i(\vec{r})}{\partial x^l}\frac{\partial \psi_i(\vec{r})}{\partial x^k}\Bigg],
\label{eq:stress}
\end{eqnarray}
where $\{k, l\} = \{1, 2, 3\}$, $m$ is the electron mass, $\psi_i(\vec{r})$ is the $i$th natural orbital and $\nu_i$ is its occupation number.

By taking a trace of $\overleftrightarrow{\tau}^{S}(\vec{r})$, we can define energy density of the quantum system at each point in space. The energy density $\varepsilon_\tau^S(\vec{r})$ is given by 
\begin{eqnarray}
\varepsilon_\tau^S(\vec{r}) = \frac{1}{2} \sum_{k=1}^3 \tau^{Skk}(\vec{r}).
\end{eqnarray}
We note that, by using the virial theorem, integration of $\varepsilon_\tau^S(\vec{r})$ over whole space gives usual total energy $E$ of the system: $\int \varepsilon_\tau^S(\vec{r}) d\vec{r} = E$.

Regional chemical potential $\mu_R$ \cite{Tachibana1999} is calculated approximately using $\varepsilon_\tau^S(\vec{r})$ \cite{Szarek2007}. 
\begin{eqnarray}
\mu_R = \frac{\partial E_R}{\partial N_R} \approx \frac{\varepsilon_\tau^S(\vec{r})}{n(\vec{r})}, \label{eq:mu}
\end{eqnarray}
where $n(\vec{r})$ is the ordinary electron density at $\vec{r}$. Since electrons tend to move from high $\mu_R$ region to low $\mu_R$ region, the distribution of $\mu_R$ maps the chemical reactivity. 

Now, we define bond orders as $\varepsilon_\tau^S(\vec{r})$ or $\mu_R$ at ``Lagrange point" \cite{Szarek2007}. The Lagrange point $\vec{r}_L$ is the point where the tension density $\vec{\tau}^S(\vec{r})$ given by the divergence of the stress tensor 
\begin{eqnarray} 
\tau^{S k}(\vec{r})&=&  \sum_l \partial_l  \tau^{Skl}(\vec{r}) \nonumber \\
&=&\frac{\hbar^2}{4m}\sum_i \nu_i
\Bigg[\psi^*_i(\vec{r})\frac{\partial \Delta\psi_i(\vec{r})}{\partial x^k}-\frac{\partial\psi^*_i(\vec{r})}{\partial x^k} \Delta\psi_i(\vec{r}) \nonumber\\
& &+\frac{\partial \Delta\psi^*_i(\vec{r})}{\partial x^k}\psi_i(\vec{r}) -\Delta \psi^*_i(\vec{r}) \frac{\partial \psi_i(\vec{r})}{\partial x^k}\Bigg],
\label{eq:tension}
\end{eqnarray}
vanishes. Namely, $\tau^{S k}(\vec{r}_L)=0$. $\vec{\tau}^S(\vec{r})$ is the expectation value of the tension density operator $\Hat{\vec{\tau}}^S(\vec{r})$, which cancels the Lorentz force density operator $\Hat{\vec{L}}(\vec{r})$ in the equation of motion for stationary state \cite{Tachibana2003}. Therefore, we see that $\vec{\tau}^S(\vec{r})$ expresses purely quantum mechanical effect and it has been proposed that this stationary point characterizes chemical bonding \cite{Szarek2007}. Then, our definitions of bond order are
\begin{eqnarray}
b_\varepsilon = \frac{\varepsilon^S_{\tau {\rm AB}}(\vec{r}_L)}{\varepsilon^S_{\tau {\rm HH}}(\vec{r}_L)}, \label{eq:be}
\end{eqnarray}
and
\begin{eqnarray}
b_\mu = \frac{\varepsilon^S_{\tau {\rm AB}}(\vec{r}_L) / n_{\rm AB}(\vec{r}_L)}{\varepsilon^S_{\tau {\rm HH}}(\vec{r}_L) / n_{\rm HH}(\vec{r}_L)}. \label{eq:bmu}
\end{eqnarray}
One should note normalization by the respective values of  a H$_2$ molecule calculated at the same level of theory (including method and basis set).

We use Molecular Regional DFT (MRDFT) package \cite{MRDFTv3} to compute these quantities introduced in this section. Some part of the visualization is done using 
PyMOL Molecular Viewer program \cite{PyMOL}.

\section{Results and discussion} \label{sec:results}

\subsection{Bond order analysis} \label{sec:bo}
\begin{figure}
\begin{center}
\includegraphics[width=14cm]{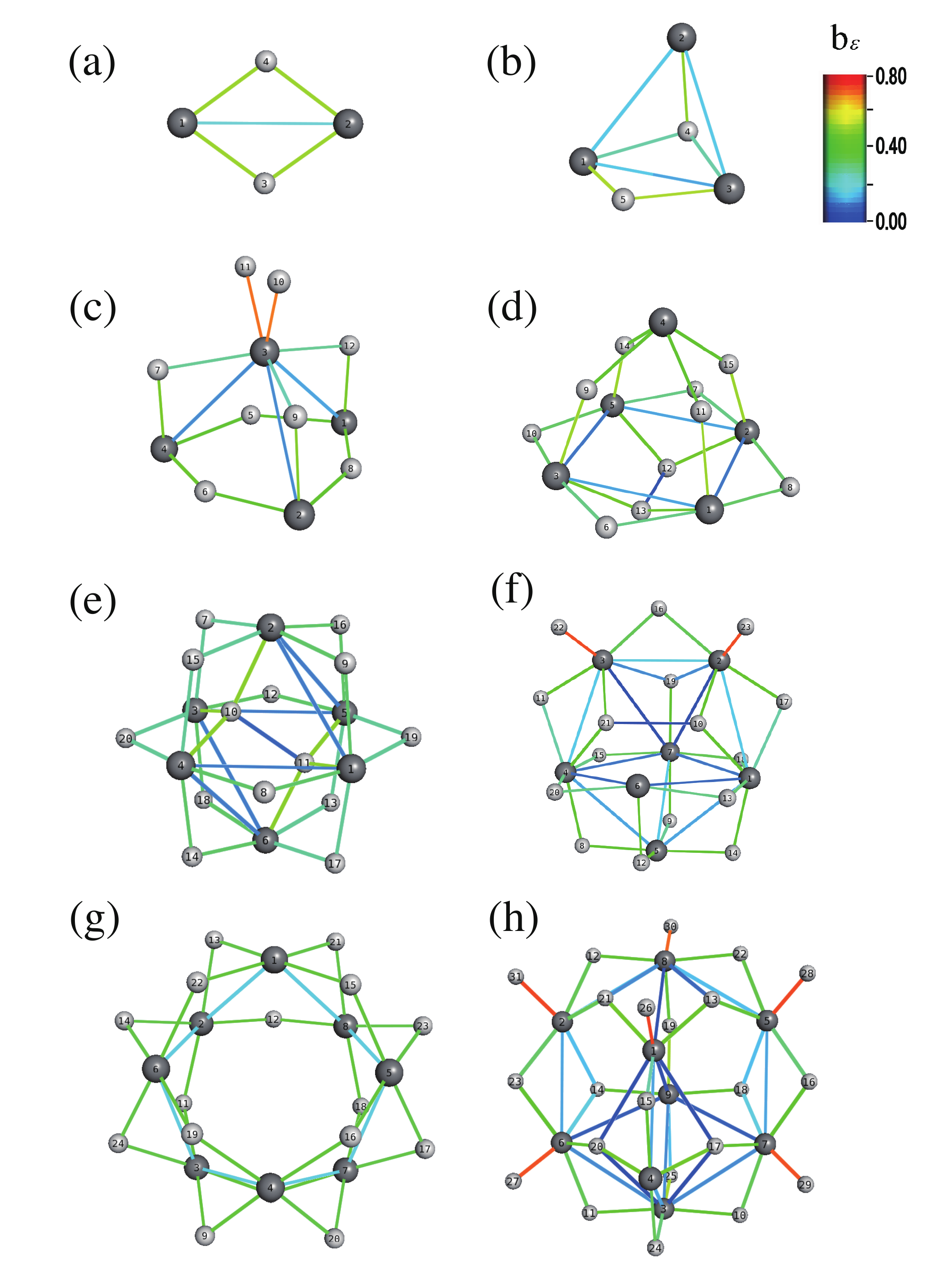}
\caption{Optimized structures and bond order for hydrogenated Pd clusters: (a) Pd$_2$H$_2$, (b) Pd$_3$H$_2$, (c) Pd$_4$H$_8$, (d) Pd$_5$H$_{10}$, (e) Pd$_6$H$_{14}$, (f)  Pd$_7$H$_{16}$, (g)  Pd$_8$H$_{16}$ and (h) Pd$_9$H$_{22}$. The bonds are drawn at which Lagrange points are found and our energy density based bond order $b_\varepsilon$  (eq.~\eqref{eq:be}) is shown by color.  }
\label{fig:be}
\end{center}
\end{figure}

\begin{figure}
\begin{center}
\includegraphics[width=14cm]{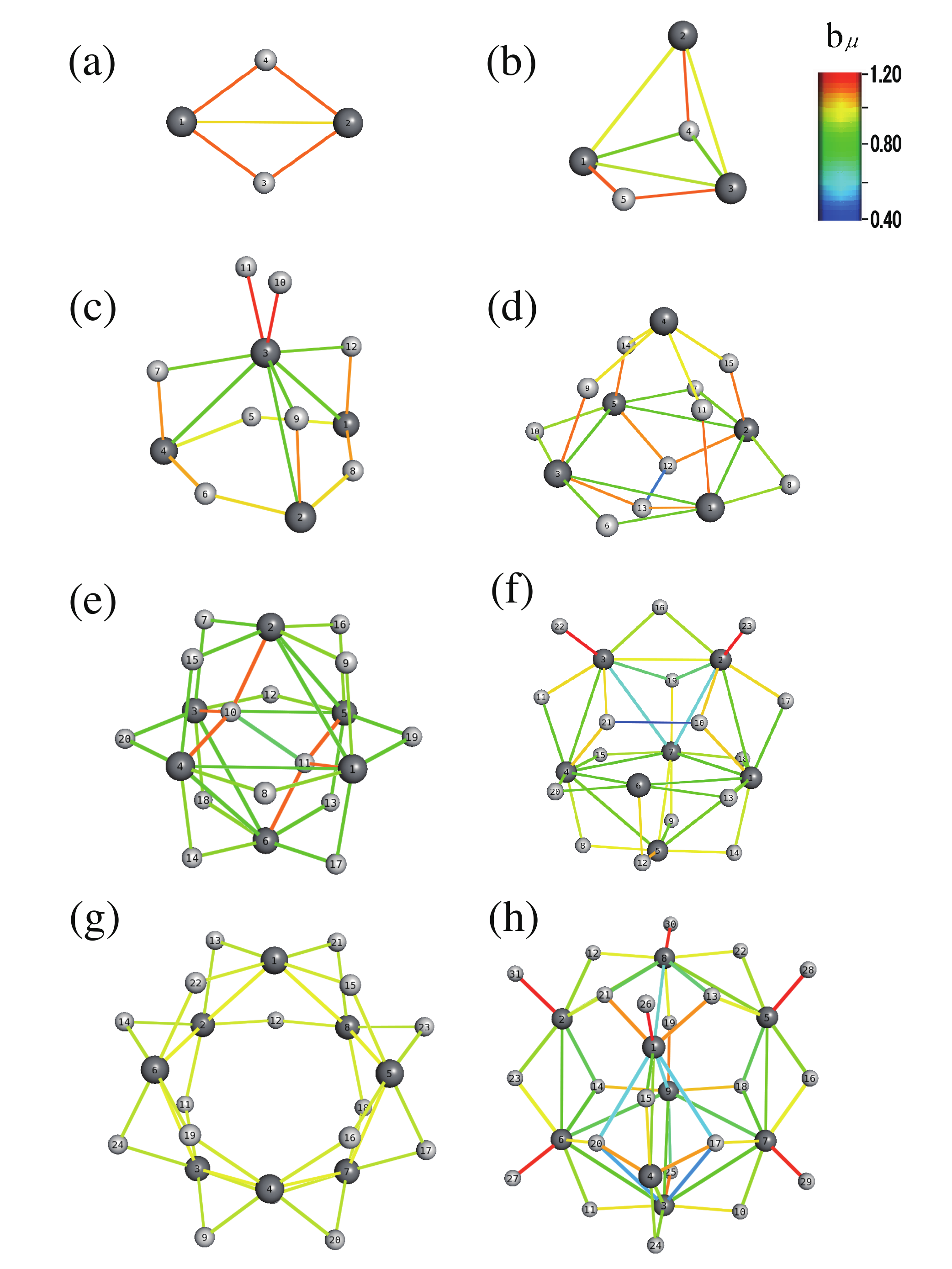}
\caption{Same as Fig.~\ref{fig:be} but with  our chemical potential based bond order $b_\mu$ (eq.~\eqref{eq:bmu})  is shown by color.  }
\label{fig:bmu}
\end{center}
\end{figure}

The optimized structures for hydrogenated Pd clusters Pd$_2$H$_2$, Pd$_3$H$_2$, Pd$_4$H$_8$, Pd$_5$H$_{10}$, Pd$_6$H$_{14}$, Pd$_7$H$_{16}$, Pd$_8$H$_{16}$ and Pd$_9$H$_{22}$ are shown in Fig.~\ref{fig:be}. In the figure, atoms are connected when the Lagrange point (Sec.~\ref{sec:RDFT}) is found between them and the bond is colored 
to show the magnitude of the bond order $b_\varepsilon$  (eq.~\eqref{eq:be}). Fig.~\ref{fig:bmu} shows the exactly the same structures with the different bond order $b_\mu$ (eq.~\eqref{eq:bmu}).
\footnote{
We list all bond orders and enlarged structures with atom numbering in the supplementary materials (Fig. S1 and Table S1).
} 
These structures are obtained by re-optimizing the structures reported in Ref.~\cite{Zhou2008}. We performed optimization with multiplicity of 1, 3 and 5 for each cluster and adopted the one with the lowest energy, which turned out to be singlet for all the clusters. We did not find much difference between the structures in Ref.~\cite{Zhou2008} and our re-optimized ones. For later use, we report that we performed similar procedure for bare Pd clusters Pd$_n$ ($n=2$--9) starting with the structures obtained in Ref.~\cite{Luo2007}. In this case triplets have the lowest energy for all the clusters. 

\begin{figure}
\begin{center}
\includegraphics[width=14cm]{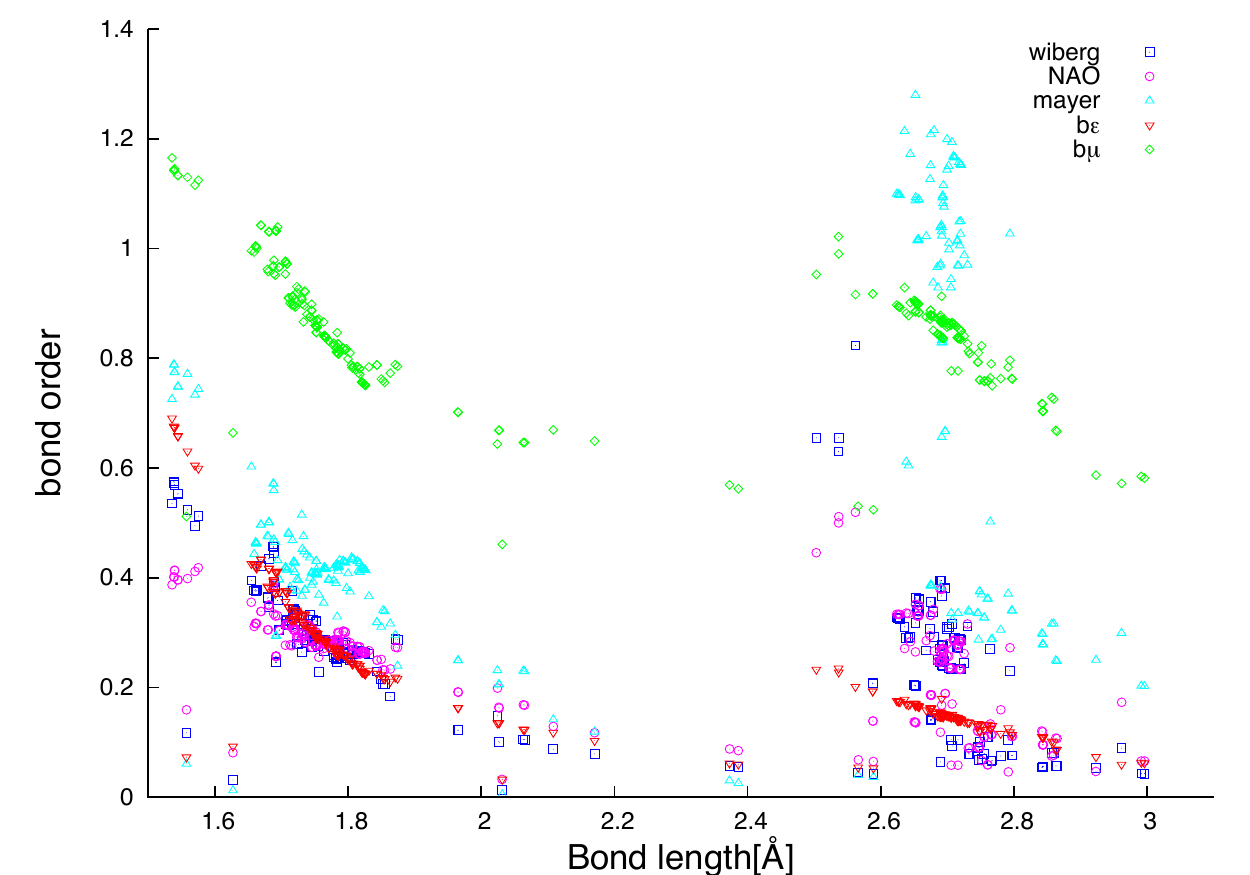}
\caption{Comparison of the relation between bond length and various bond orders: the Wiberg bond index (blue square), atom-atom overlap-weighted NAO bond order (magenta circle), the Mayer's bond order (light-blue triangle) and our bond orders $b_\varepsilon$ (red down-triangle) and $b_\mu$ (green diamond). }
\label{fig:dist_bo_compare}
\end{center}
\end{figure}

To show features of the chemical bonds in those clusters collectively and to exhibit usefulness of our bond order definitions, we plot bond orders against the bond length for all the bonds in the bare and hydrogenated Pd clusters in Fig.~\ref{fig:dist_bo_compare}. In addition to our bond orders $b_\varepsilon$ and $b_\mu$, we plot using conventional bond orders: the Wiberg bond index \cite{WibergBO}, atom-atom overlap-weighted natural atomic orbital (NAO) bond order \cite{NAOBO1,NAOBO2} and the Mayer's bond order \cite{MayerBO}. It is apparent that our bond orders have better correlation than the other conventional bond orders. Therefore, we only show our bond orders for the following analysis.  

\begin{figure}
\begin{center}
\includegraphics[width=10cm]{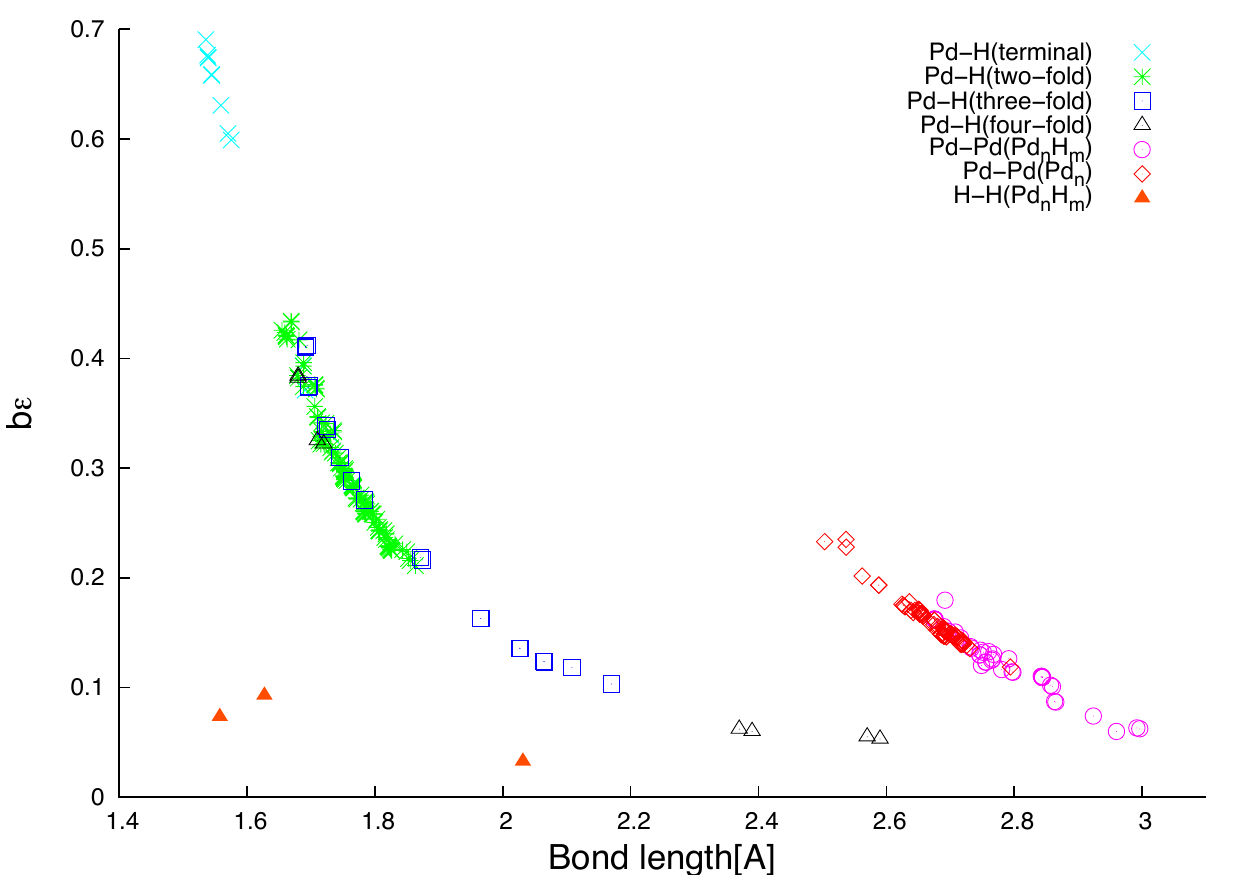}
\caption{The relation between bond length and energy density bond order $b_\varepsilon$ in the bare Pd clusters and hydrogenated Pd clusters. We classify bonding types as terminal Pd--H (light-blue cross), two-fold Pd-H (green asterisk), three-fold Pd--H (blue square), four-fold Pd--H (black triangle), Pd--Pd in hydrogenated Pd clusters (magenta circle), Pd--Pd in bare Pd clusters (red diamond), and H--H (red filled-triangle). Note that the H$_2$ molecule would locate at (0.748, 1.000). }
\label{fig:dist_be}
\end{center}
\end{figure}

\begin{figure}
\begin{center}
\includegraphics[width=10cm]{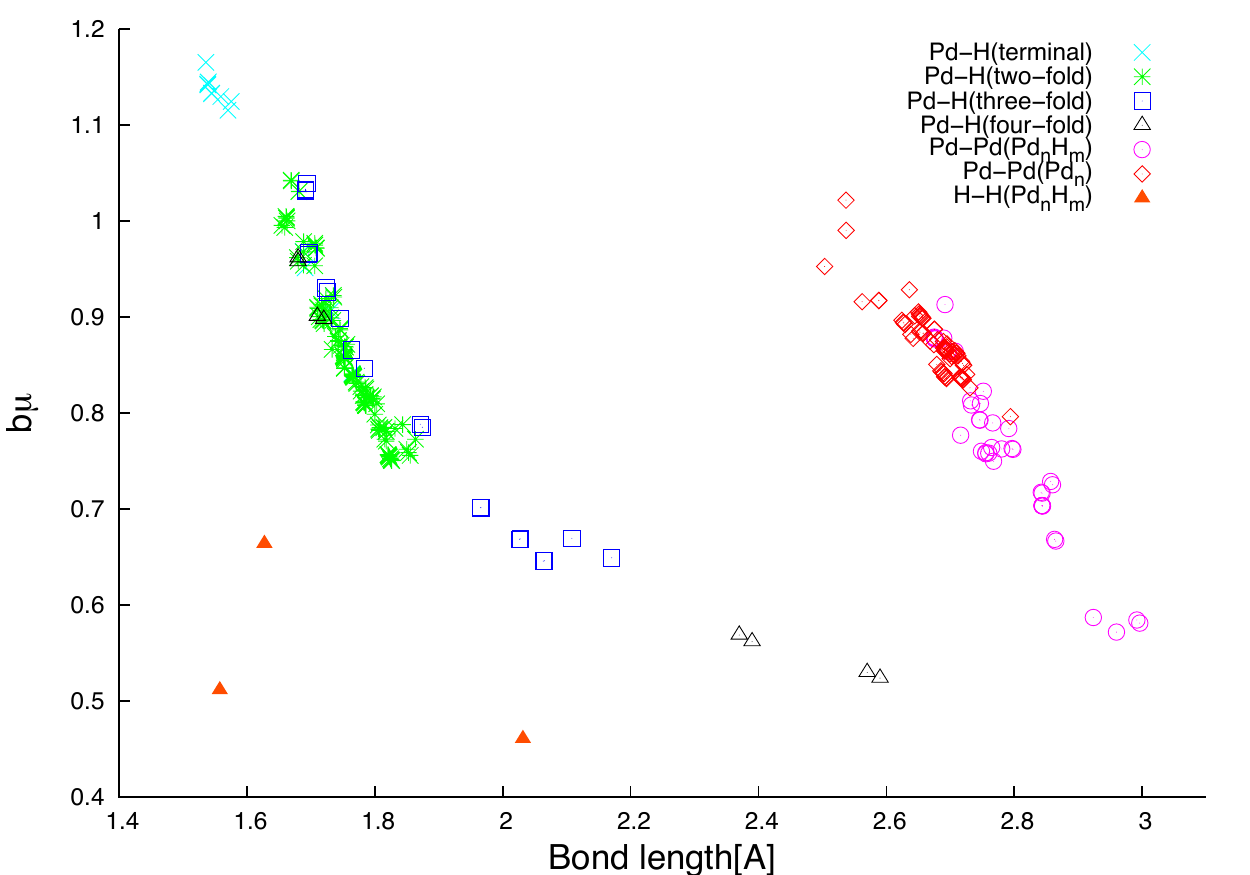}
\caption{Same as Fig.~\ref{fig:dist_be} but with chemical potential bond order $b_\mu$.}
\label{fig:dist_bmu}
\end{center}
\end{figure}

In Figs.~\ref{fig:dist_be} and \ref{fig:dist_bmu}, we re-plot the bond order v.s. bond length for $b_\varepsilon$ and $b_\mu$ respectively, this time distinguishing between different types of bonding. The types we consider are Pd--Pd, terminal Pd--H, two-fold Pd-H, three-fold Pd--H, four-fold Pd--H and H--H. Here, two-fold Pd-H bond is associated with the H atom bridging two Pd atoms like Pd--H--Pd and, similarly, three-(four-)fold Pd-H bond with the H atom bonded to three (four) Pd atoms. Each type of bonding is plotted by different marks in Figs.~\ref{fig:dist_be} and \ref{fig:dist_bmu} and number of each type for each cluster is shown in Table \ref{tab:bond_type}.
We note that some of the H atoms which we classified as having two-fold and three-fold Pd--H bonds have a bond with H atom in addition. In detail, H(12) and H(13) in Pd$_5$H$_{10}$ and H(10) and H(21) in Pd$_7$H$_{16}$ are bonded to each other in addition to two Pd atoms, which results in having three bonds from each H. Similarly, H(10) and H(11) in Pd$_6$H$_{14}$ are bonded to each other in addition to three Pd atoms, making four bonds from each H.

From Figs.~\ref{fig:dist_be} and \ref{fig:dist_bmu}, we see that bonds in the Pd clusters can be classified by the different slopes on the bond order v.s. bond length plane. There are a slope that corresponds to the Pd--Pd bonds and two slopes for Pd--H bonds. There are three outliers that correspond to H--H bonds. The fact that Pd--Pd bonds, whether they are in hydrogenated clusters or in bare clusters, are on a single slope indicates that the character of the bonding is not affected much by the hydrogenation. They on average become longer and weaker upon hydrogenation but the relation between the bond order and bond length is unchanged. 

 Pd--H bonds may be classified in two groups. One has shorter bond length ($\lesssim 1.9\,{\rm \AA}$) and higher inclination to which terminal Pd--H bond and two-fold Pd--H belong. Some bonds in three-fold and four-fold Pd--H also belong to this group. Another group has longer bond length ($\gtrsim 1.9\,{\rm \AA}$) and lower inclination and consists only of three-fold and four-fold Pd--H. For convenience, we call the former group ``A" and the latter ``B". Closer inspection of these H atoms with high-coordination numbers tells that there is no H atom which has only group B bonds. The bonds stem from H atoms consist of those of group A or mixture of A and B. See Table \ref{tab:bond_type_2} for the detail. It is interesting that there are Pd--H bonds which are longer than some of the Pd--Pd bonds. Such long bonds are found in four-fold Pd--H bonds, which will be investigated more in Sec.~\ref{sec:stress}. 
We also discuss in Sec.~\ref{sec:stress} that group A is characterized by a ``spindle structure" and the group B by a ``pseudo-spindle structure".

We now would like to give somewhat more quantitative analysis of these slopes in the bond order v.s.~bond length relation by fitting the data points to linear and exponential curves.
We fit data points which belong to the shorter Pd--H bonds (group A), the longer Pd--H bonds (group B) and the Pd--Pd bonds to functions in the forms
$y = a x + b$ and $y = c \exp (-d x)$. 
Here, each group has 169, 13 and 134 data points, and in the fitting functions, $y$ stands for $b_\varepsilon$ or $b_\mu$ and $x$ stands for the bond length.
We summarize the results in Table \ref{tab:fit}.
Roughly speaking, both functional forms give good fits, and the differences in the fitting parameters  confirm the existence of two slopes for the Pd--H bonds (e.g. the inclination of the linear fits, the parameter $a$, is about 8 times larger for the group A than the group B).
On closer look, the fits to $b_\varepsilon$ are better performed by the exponential function especially for Pd--H bonds.
 In contrast, the use of exponential form does not improve the fits to $b_\mu$ much, and in fact, the linear fits are slightly better for the Pd--H bond group A and the Pd--Pd bond group.

As mentioned above, H--H bonds are shown in Figs.~\ref{fig:dist_be} and \ref{fig:dist_bmu} as isolated points from the slopes of Pd--Pd and Pd--H. They cannot be put on a single slope and appear somewhat irregular. Since the hydrogenated Pd clusters here are formed by dissociative chemisorption of H$_2$ \cite{Zhou2008}, these H--H are considered to be not trivial bonding of the H$_2$ molecule. Namely, they are formed on or within the clusters and characteristic to the Pd clusters. They all have much longer and weaker bonds than the H--H bond in the H$_2$ molecule.
(Note that $b_\varepsilon$ and $b_\mu$ are unity for the H$_2$ molecule by definition. The bond length of H$_2$ in our computational method is 0.748252\,\AA.) 
We will investigate these H--H bonds in more detail in Sec.~\ref{sec:stress}. The existence of such H--H bonds are most notable difference from hydrogenated small Pt clusters which have been investigated in Refs.~\cite{Zhou2007, Szarek2009a}

\subsection{Stress tensor analysis of chemical bond} \label{sec:stress}
In the previous section, we have seen that there are several interesting bonding patterns found in the hydrogenated Pd clusters. We investigate these bonds in detail via the electronic stress tensor analysis. 

\begin{figure}
\begin{center}
\includegraphics[width=17cm]{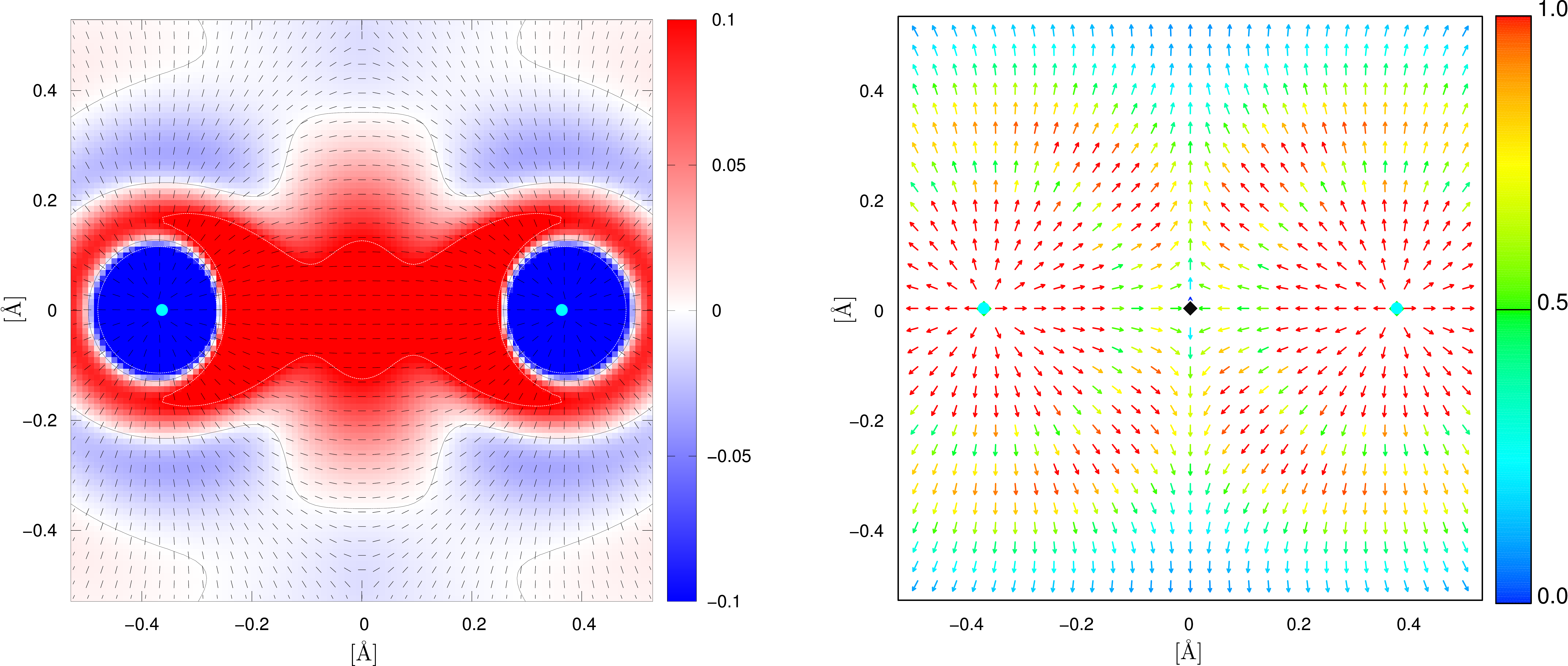}
\caption{The H--H bond in H$_2$. The largest eigenvalue of the stress tensor and corresponding eigenvector (left panel) and tension (right panel) are plotted on the plane including the bond axis. On the left panel, contours of 0.1 and $-0.1$ are shown by white dotted lines. On the right panel, the normalized tension vectors whose norm is expressed by the color of the arrows are shown and the Lagrange point is marked by a black diamond.}
\label{fig:H2_HH}
\end{center}
\end{figure}

Before we discuss the Pd clusters, we show how the chemical bond of the hydrogen molecule is expressed by the electronic stress tensor (Eq.~\eqref{eq:stress}). In Fig.~\ref{fig:H2_HH}, on the left panel, we plot the largest eigenvalue of the stress tensor and corresponding eigenvector on the plane including the internuclear axis. On the right panel, we plot the tension vector (Eq.~\eqref{eq:tension}), which is normalized and whose norm is expressed by the color of the arrows. The sign of the largest eigenvalue tells whether electrons at a certain point in space feel tensile force (positive eigenvalue) or compressive force (negative eigenvalue) and the eigenvector tells the direction of the force. We can see that the region with positive eigenvalue spreads between the H atoms, which corresponds to the formation of a covalent bond. In that region, the eigenvectors form a bundle of flow lines that connects the H nuclei. Such a region, called ``spindle structure" \cite{Tachibana2004}, is clearly seen in the panel. On the right panel, the vanishing point of tension vector ``Lagrange point" (Sec.~\ref{sec:RDFT}) is found at the midpoint of the internuclear axis, which is quite reasonable. 

\begin{figure}
\begin{center}
\includegraphics[width=17cm]{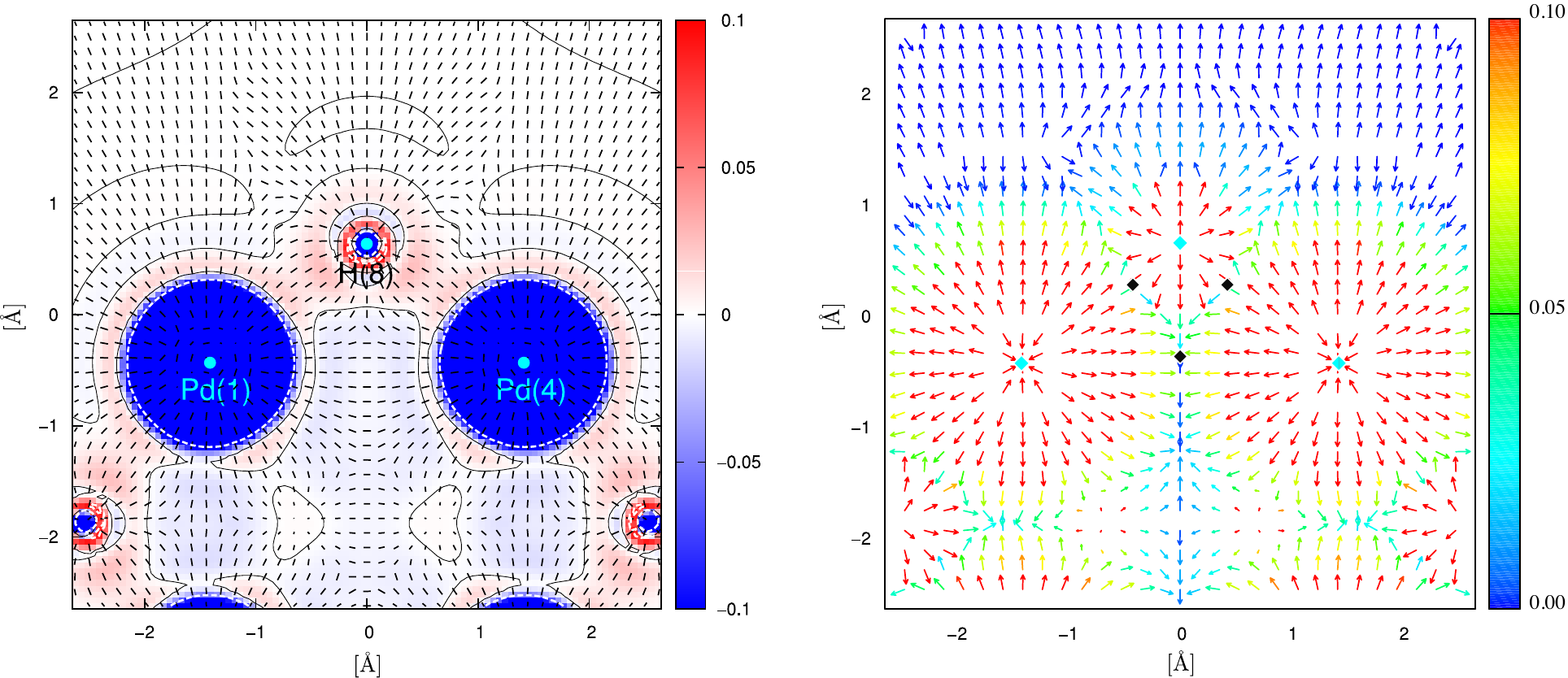}
\caption{The Pd--H--Pd bridging bond in Pd$_6$H$_{14}$ is shown as the largest eigenvalue of the stress tensor and corresponding eigenvector (left panel) and tension (right panel) on the plane including the labelled atoms. See Figs.~\ref{fig:be} or \ref{fig:bmu} for the number in the label. On the right panel, the Lagrange point is marked by a black diamond.}
\label{fig:Pd6H14_bridge}
\end{center}
\end{figure}

We will now turn to the hydrogenated Pd clusters. We first look at the Pd--H--Pd bridging bond in Pd$_6$H$_{14}$ as shown in Fig.~\ref{fig:Pd6H14_bridge}. We see the spindle structure between Pd and H just like the H$_2$ molecule mentioned above, indicating the covalency of the Pd--H bond here. As for the Pd--Pd bond, although we found a Lagrange point and flow of the eigenvectors connecting Pd atoms, the eigenvalue in the region between Pd atoms has negative value. This indicates that the interaction between these two Pd atoms is weak and the bond between them is different from a covalent bond 
characterized by a spindle structure. We shall call this pattern a ``pseudo-spindle structure". Namely, the pseudo-spindle structure has similar eigenvector flow to that of the spindle structure between two atoms but with negative eigenvalue region between them. 

\begin{figure}
\begin{center}
\includegraphics[width=17cm]{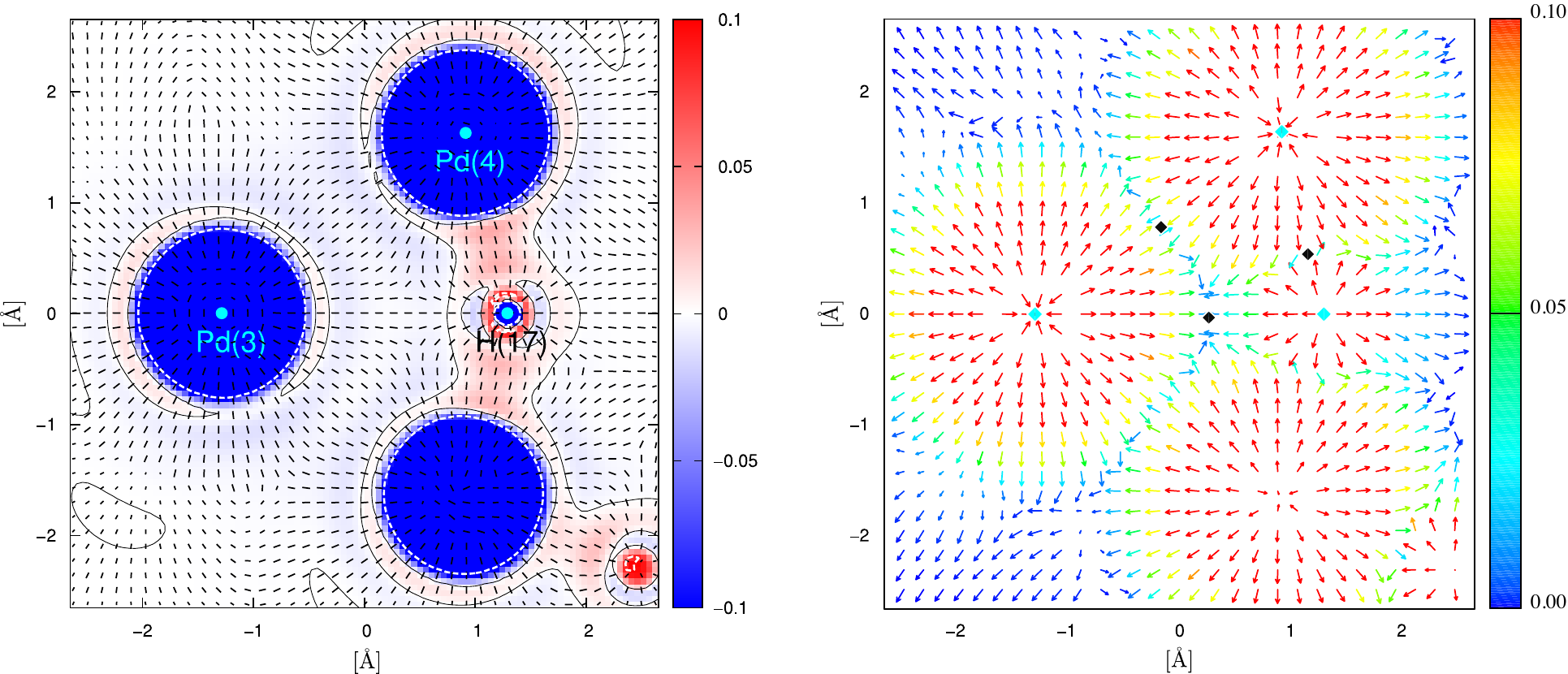}
\caption{The four-fold Pd--H bond in Pd$_9$H$_{22}$ is shown in the similar manner as Fig.~\ref{fig:Pd6H14_bridge}. Two of four Pd--H bonds from H(17) are shown on this plane. }
\label{fig:Pd9H22_PdH}
\end{center}
\end{figure}

We next examine the very weak Pd--H bond which is found in the four-fold bond in Pd$_9$H$_{22}$. There are two H atoms which have bonds between four Pd atoms and two of them are very weak (see Tables \ref{tab:bond_type} and \ref{tab:bond_type_2}). Fig.~\ref{fig:Pd9H22_PdH} focuses on one of such H atom (H(17)) and shows one weaker Pd--H bond (Pd(3)--H(17)) and one stronger bond (Pd(4)--H(17)) (other two bonds, Pd(1)--H(17) and Pd(7)--H(17), are off this plane). The stronger one has the spindle structure and is similar to the one in the Pd--H--Pd bridging bond. Although the weaker one has a long bond distance (2.59\,${\rm \AA}$) and negative eigenvalue region around the Lagrange point, there is a flow which connects Pd(3) and H(17) so we regard this as a bond. In other words, this bond is characterized by the pseudo-spindle structure.
This classification that shorter bonds are characterized by spindle structures and longer ones are by pseudo-spindle structures seems to hold rather in general for the hydrogenated Pd clusters we have investigated. In Sec.~\ref{sec:bo}, we have defined two groups ``A" and ``B" for the Pd--H bonds where the former (the latter) has bond length shorter (longer) than about 1.9\,\AA. In Sec.~\ref{sec:bo}, we have pointed out that they are on slopes with different inclination in the bond order v.s. bond length plot. From the viewpoint of the electronic stress tensor, group A is characterized by a spindle structure and the group B by a pseudo-spindle structure.

\begin{figure}
\begin{center}
\includegraphics[width=17cm]{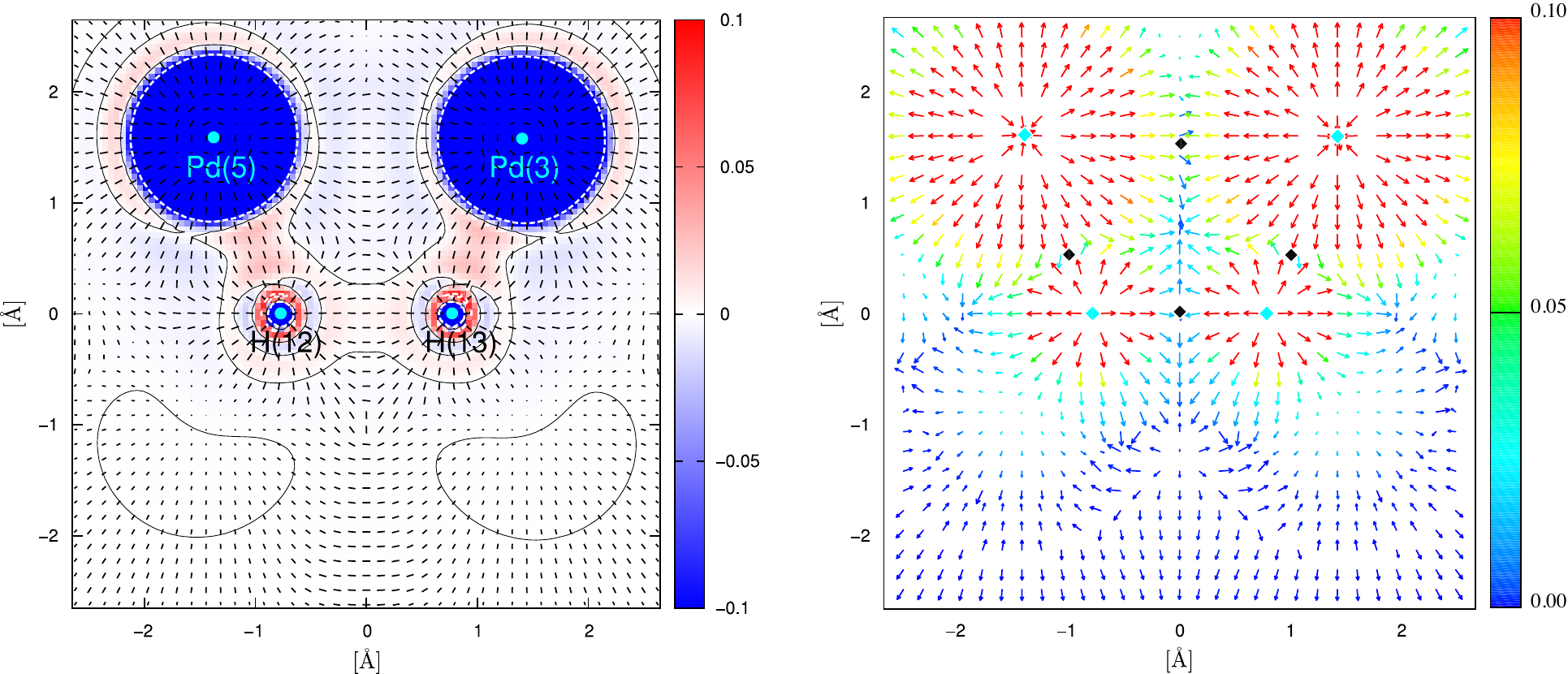}
\caption{The H--H bond in Pd$_5$H$_{10}$ is shown in the similar manner as Fig.~\ref{fig:Pd6H14_bridge}.}
\label{fig:Pd5H10_HH}
\end{center}
\end{figure}

\begin{figure}
\begin{center}
\includegraphics[width=17cm]{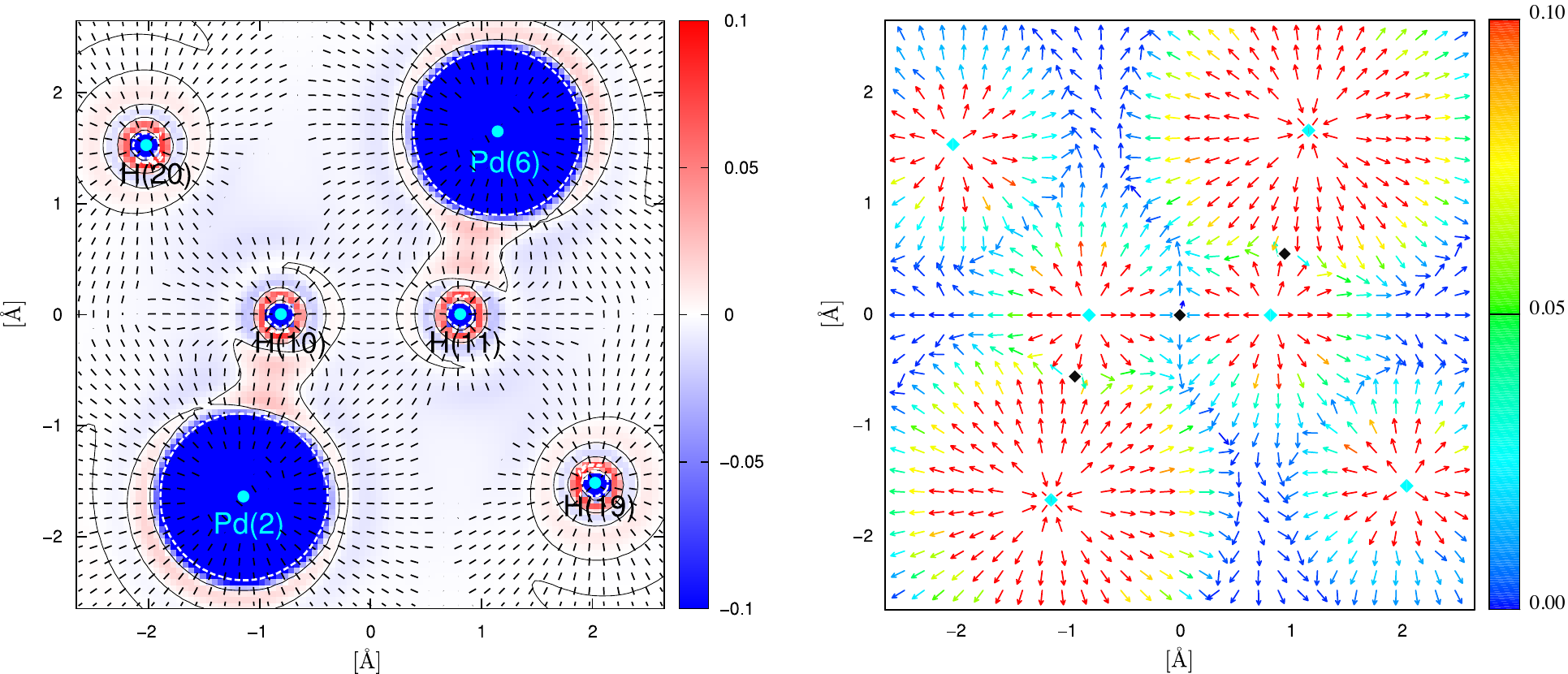}
\caption{The H--H bond in Pd$_6$H$_{14}$ is shown in the similar manner as Fig.~\ref{fig:Pd6H14_bridge}.}
\label{fig:Pd6H14_HH}
\end{center}
\end{figure}

\begin{figure}
\begin{center}
\includegraphics[width=17cm]{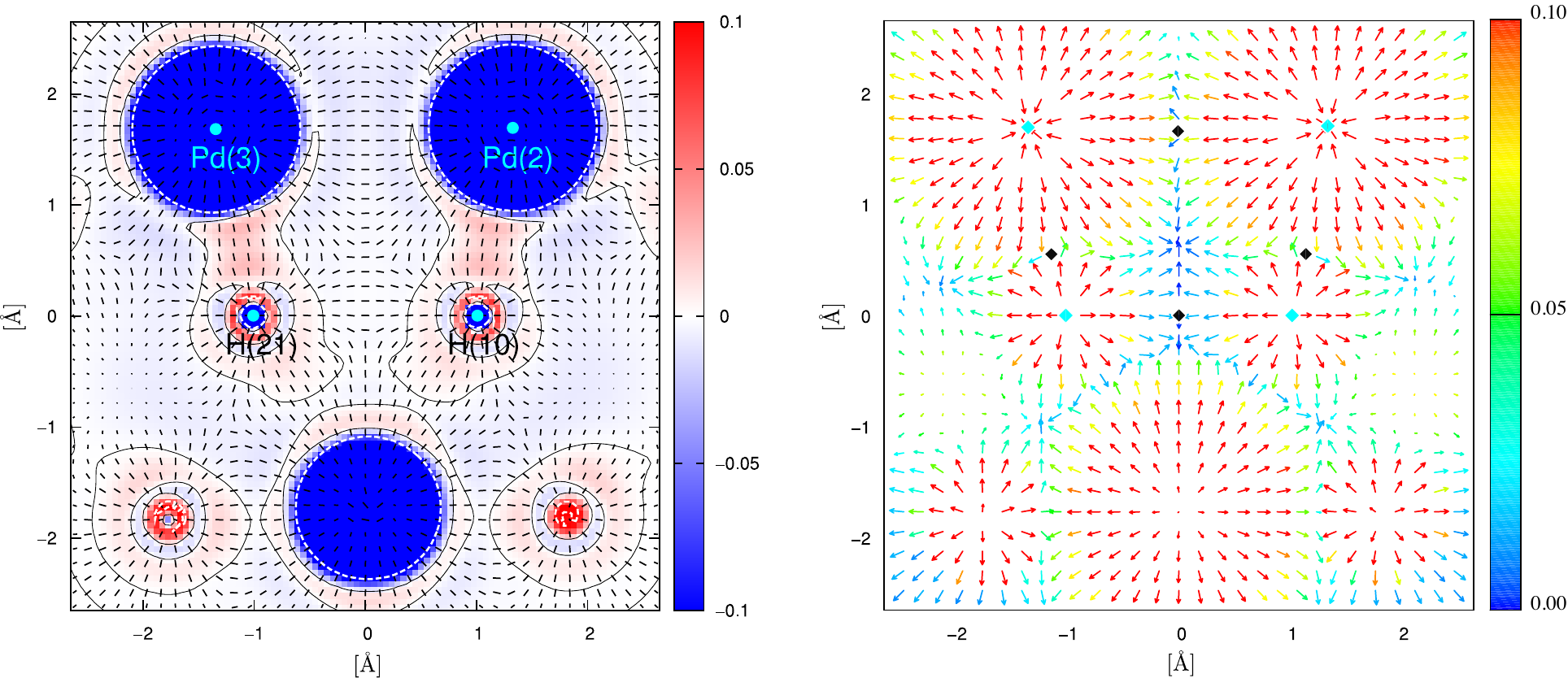}
\caption{The H--H bond in Pd$_7$H$_{16}$ is shown in the similar manner as Fig.~\ref{fig:Pd6H14_bridge}.}
\label{fig:Pd7H16_HH}
\end{center}
\end{figure}

Finally, we discuss the H--H bonds in Pd$_5$H$_{10}$, Pd$_6$H$_{14}$ and Pd$_7$H$_{16}$. Their bond lengths are respectively 1.56\,${\rm \AA}$, 1.63\,${\rm \AA}$ and 2.03\,${\rm \AA}$, which are longer than that of the H$_2$ molecule. From the structural point of view, the H--H in Pd$_5$H$_{10}$ is located outside the Pd cage. The H--H in Pd$_6$H$_{14}$ is completely contained in the Pd cage and the one in Pd$_7$H$_{16}$ is marginally within the Pd cage. As before, we show the electronic stress tensor and tension for these bonds in Figs.~\ref{fig:Pd5H10_HH}, \ref{fig:Pd6H14_HH} and \ref{fig:Pd7H16_HH}. In these figures, there are more than three labelled atoms but they are all on the same plane (the distance between the plane and the labelled atoms are less than 0.01\,${\rm \AA}$). In Fig.~\ref{fig:Pd7H16_HH}, left panel, there is a region with negative eigenvalue (shown by a blue circular region) at the lower-center part of the panel. This is caused by the existence of Pd(6) but, since it locates slightly above this plane (about 0.4\,${\rm \AA}$), it is not labelled. 

In these figures, we see that all the regions between Pd and H have a spindle structures, suggesting covalency of the Pd--H bonds. As for the H--H bonds, there is a spindle structure for the one in Pd$_5$H$_{10}$ (Fig.~\ref{fig:Pd5H10_HH}).  The H--H bonds in Pd$_6$H$_{14}$ and Pd$_7$H$_{16}$ (Figs.~\ref{fig:Pd6H14_HH} and \ref{fig:Pd7H16_HH}) are characterized by the negative eigenvalue region and eigenvectors connecting H atoms, namely a pseudo-spindle structure. 

\subsection{Improving bond order definition by surface integral} \label{sec:beS}
In this section, we propose a way to improve our definition of bond order and apply it to the hydrogenated Pd clusters. 
As is described in Sec.~\ref{sec:RDFT}, our bond order $b_\varepsilon$ is defined from the energy density evaluated at the Lagrange point. 
In the previous paper  \cite{Szarek2008}, using many kinds of hydrocarbon molecules, it has been shown that this definition manifests a nice feature as a bond order.

However, it is not difficult to imagine a type of chemical bond which cannot be well characterized by a single point between two atoms.
This would be true for chemical bonds where spatially extended $d$-orbitals are involved. Then, we consider it is worthwhile to investigate
this issue by expanding our bond order definition by taking the surface integral of energy density instead of the energy density at the Lagrange point. 

Then, we need to determine the surface over which we integrate the energy density. 
The most natural choice would be a ``Lagrange surface" \cite{Tachibana2010} which is constructed from a family of lines which going out from a Lagrange point (if a Lagrange surface includes a Lagrange point).
Namely, we define bond order of the bond between atoms A and B as
\begin{eqnarray}
b_{\varepsilon(S)} =
\frac{ \int_{{\cal S}_{\rm AB}} d^2\sigma \varepsilon^S_\tau(\vec{\sigma})}{ \int_{{\cal S}_{\rm HH}} d^2\sigma \varepsilon^S_\tau(\vec{\sigma})},
\end{eqnarray}
where ${\cal S}_{\rm AB}$ denotes the Lagrange surface between atoms A and B.
As is the cases of $b_\varepsilon$ (Eq.~\eqref{eq:be}) and $b_\mu$ (Eq.~\eqref{eq:bmu}), we normalize by the value of hydrogen molecule. 

Unfortunately, however, this Lagrange surface is not so easy to define numerically. 
Hence we instead take the surface integral over the plane which includes a Lagrange point and is perpendicular to the axis connecting two atoms.
Note that such a plane coincides with a Lagrange surface in the case of homonuclear diatomic molecules. 

\begin{figure}
\begin{center}
\includegraphics[width=12cm]{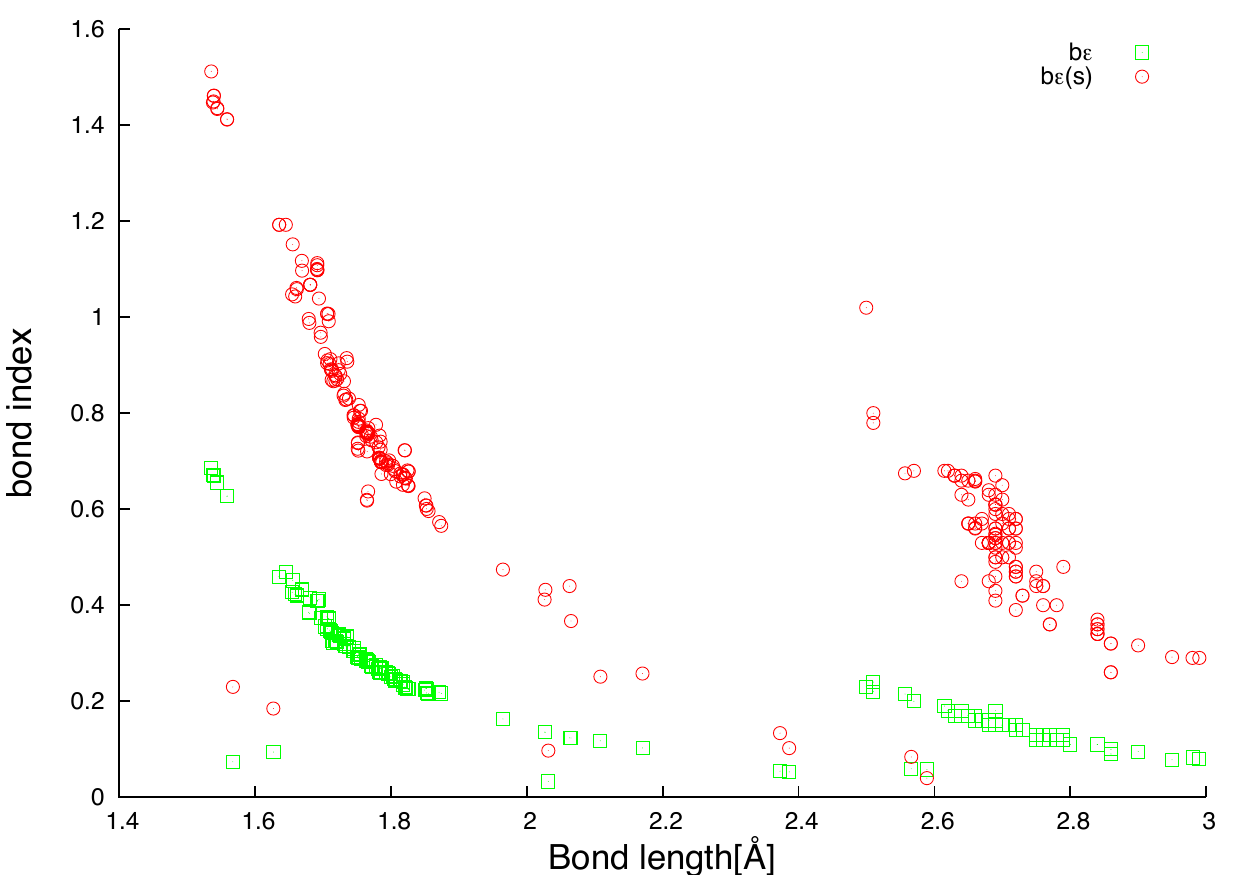}
\caption{Comparison of the relation between bond length and bond orders. Bond order defined from the energy density at the Lagrange points ($b_\varepsilon$) and 
one defined from surface integral of the energy density ($b_{\varepsilon(S)}$). }
\label{fig:beS}
\end{center}
\end{figure}

\begin{figure}
\begin{center}
\includegraphics[width=12cm]{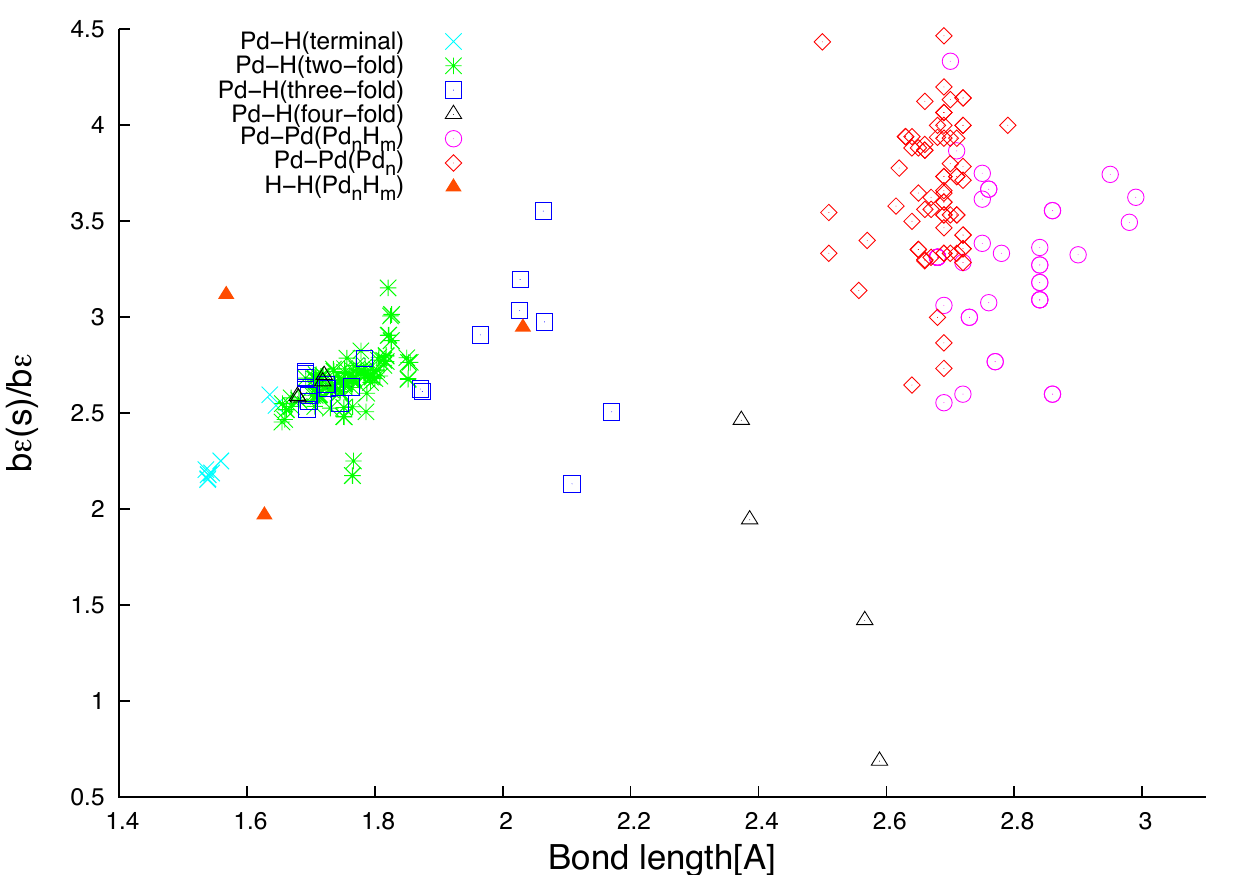}
\caption{The ratio of $b_{\varepsilon(S)}$ to $b_\varepsilon$. }
\label{fig:beS_be}
\end{center}
\end{figure}

The results of the $b_{\varepsilon(S)}$ calculation are shown in Figs.~\ref{fig:beS} and \ref{fig:beS_be}.
We see that $b_{\varepsilon(S)}$ is calculated to be larger than $b_{\varepsilon}$ for most of the bonds in the hydrogenated Pd clusters. 
This comes from the fact that the energy density distribution in those clusters is spatially extended relative to that of the hydrogen molecule. 
The ratio of $b_{\varepsilon(S)}$ to $b_\varepsilon$ is especially large for Pd--Pd bonds as shown in Fig.~\ref{fig:beS_be}.
This in turn is considered to be due to the $d$-orbitals of Pd atoms participating in the bonds.

\section{Summary} \label{sec:summary}
In this paper, we have applied recently developed method to analyze electronic structure via electronic stress tensor to the small hydrogenated Pd clusters whose 
structures have been known \cite{Luo2007,Zhou2008}. From the results of ab initio electronic structure calculation of these clusters, we calculated quantities which are defined at each point
in space: stress tensor, tension, energy density and regional chemical potential. The chemical bond is characterized by the Lagrange point, where the tension
vanishes and bond orders are defined by the energy density or chemical potential at that point. We have confirmed that thus defined bond orders are useful to classify the chemical bonds in Pd clusters as had been done for Pt clusters \cite{Szarek2009a} and Al$_4$ clusters \cite{Ichikawa2010b}. 

For some of the bonds, we have done more detailed analysis by drawing the eigenvalue of stress tensor, corresponding eigenvectors and tension vectors. In particular, the weak bonds with long bond distances suggested by the Lagrange point search in our bond order analysis are confirmed to have a flow of eigenvectors connecting bonded atoms. 
They include four-fold Pd--H bonds and non-trivial H--H bonds. As for the H--H bond, we found that the one in Pd$_5$H$_{10}$ has a positive eigenvalue region, ``spindle structure", showing covalency like that of the H$_2$ molecule whereas those in Pd$_6$H$_{14}$ and Pd$_7$H$_{16}$ have negative eigenvalue regions, ``pseudo-spindle structure".
As for the Pd--H bond, it has been shown to be classified into two groups.
One with shorter bond lengths which is characterized by a spindle structure and another with longer bond lengths characterized by a pseudo-spindle structure. 

We proposed a terminology ``pseudo-spindle structure" in this paper but actually, there already has been such a structure found in our previous study for C$_2$H$_2$ \cite{Tachibana2005}. 
The negative eigenvalue of C$_2$H$_2$ is caused by the compressive stress nearby the C nuclei. 
In general, the stress tensor has a large negative eigenvalue in radial direction in neighborhood of a nucleus due the dominance of the attractive Coulomb force. 
In the case of C$_2$H$_2$, the bond length is too short that the internuclear region is immersed under the atomic compressive stress \cite{Tachibana2005}. 
This is a pseudo-spindle structure in a strong bond.
We can say that we have found in this paper two more types of pseudo spindle structure. 
One is the pseudo-spindle structure associated with very long and weak H--H bond and Pd--H bond. 
Another is the pseudo-spindle structure associated with a bond between metallic atoms, Pd--Pd bond.

We also have introduced an extension to our bond order. The modified definition uses the integration of the energy density over the ``Lagrange surface" instead of the energy density at the Lagrange point. This modification makes the bond order greater, which reflects the contribution of spatially extended $d$-orbitals to the bonds in the clusters. 

We believe that this study has provided another useful example of our stress tensor approach to chemical bonds. Further applications to other compounds, especially to those including transition metals, will solidify the basis of the stress tensor analysis and will deepen our understanding of chemical bonds.

\noindent 
\section*{Acknowledgments}
Theoretical calculations were partly performed using Research Center for Computational Science, Okazaki, Japan.
This research work is supported partly by Collaborative Research Program for Young Scientists of ACCMS and IIMC, Kyoto University (K.~I.)
The work conducted at CUG was supported by the National Natural Science Foundation of China, Grant No. 20973159 (C.~Z.).
This work is partially supported by Grant-in-Aid for Scientific research from the Ministry of
Education, Science, Sports, and Culture, Japan, No. 22550011 (A.~T.).



\newpage

\begin{table}
\caption{Number of bonding type in the hydrogenated Pd clusters. Pd--H (terminal), two-fold Pd--H (bridging), three-fold Pd--H and four-fold Pd--H respectively count the number of H atoms who have one, two, three and four bonds between Pd atom.}
\begin{center}
\begin{tabular}{|c|c|c|c|c|c|c|}
\hline
Cluster & Pd--Pd & Pd--H        & Two-fold & Three-fold & Four-fold & H--H \\
              &               & (terminal) &  Pd--H     & Pd--H       &    Pd--H &       \\
\hline
\hline
Pd$_2$H$_2$ & 1 & 0 & 2 & 0 & 0& 0 \\
Pd$_3$H$_2$ & 3 & 0 & 1 & 1 & 0& 0 \\
Pd$_4$H$_8$ & 3 & 2 & 6 & 0 & 0 & 0 \\
Pd$_5$H$_{10}$ & 4 & 0 & 10 & 0 & 0& 1 \\
Pd$_6$H$_{14}$ & 6 & 0 & 12 & 2 & 0 & 1 \\
Pd$_7$H$_{16}$ & 12 & 2 & 13 & 1 & 0& 1 \\
Pd$_8$H$_{16}$ & 8 & 0 & 16 & 0 & 0 & 0\\
Pd$_9$H$_{22}$ & 13 & 6 & 10 & 4 & 2 & 0 \\
\hline
\end{tabular}
\end{center}
\label{tab:bond_type}
\end{table}%

\begin{table}
\caption{The bond length of the Pd--H bonds participating in three-fold and four-fold Pd--H. See Fig.~\ref{fig:be} for the label of H atom. }
\begin{center}
\begin{tabular}{|c|c|l|}
\hline
Cluster & Bond center & Bond lengths [${\rm \AA}$]    \\
\hline
\hline
Pd$_3$H$_2$ & H(4) & 1.69, 1.87, 1.87 \\
\hline
Pd$_6$H$_{14}$ & H(10) & 1.69, 1.69, 1.69 \\
 & H(11) & 1.69, 1.69, 1.69 \\
 \hline
Pd$_7$H$_{16}$ & H(19) & 1.75, 2.06, 2.06 \\
\hline
Pd$_9$H$_{22}$ & H(13) & 1.69, 1.76, 2.17 \\
 & H(14) & 1.72, 1.97, 2.03 \\
 & H(17) & 1.68, 1.71, 2.39, 2.59 \\
 & H(18) & 1.73, 1.97, 2.03 \\
 & H(20) & 1.68, 1.72, 2.37, 2.57 \\
 & H(21) & 1.70, 1.78, 2.11 \\
\hline
\end{tabular}
\end{center}
\label{tab:bond_type_2}
\end{table}%

\begin{table}
\caption{Linear and exponential fits to the  bond order v.s.~bond length data. 
We compute fits to three groups of data points separately: shorter Pd--H bonds (group A),  longer Pd--H bonds (group B) and Pd--Pd bonds.
In the fitting functions, $x$ stands for the bond length. $\chi^2_{\rm red}$ stands for the reduced $\chi^2$ of the fit. 
}
\begin{center}
\begin{tabular}{|c||c|c||c|c|}
\hline
Bond Types &  $ b_\varepsilon = a x + b$  & $b_\varepsilon  = c \exp (-d x)$ & $ b_\mu = a x + b$ & $b_\mu  = c \exp (-d x)$  \\
\hline
\hline
Pd--H (A) & $a = -1.43$                                                 & $c = 2.42 \times 10^2$                            & $a = -1.40$                                                     &  $c = 11.5$  \\
                 &  $b = 2.81$                                                   & $d=3.82$                                                      & $b = 3.33$                                                    &   $d = 1.47$  \\
                  & $\chi^2_{\rm red}=5.9\times 10^{-4}$  &$\chi^2_{\rm red}=1.3\times 10^{-4}$  &$\chi^2_{\rm red}=7.3\times 10^{-4}$  &$\chi^2_{\rm red}=8.3\times 10^{-4}$  \\
\hline
Pd--H (B) & $a = -0.173$                                                 & $c = 8.53$                                                  & $a = -0.270$                                                     &  $c = 1.66$  \\
                 &  $b = 0.487$                                                   & $d = 2.04$                                                      & $b = 1.22$                                                    &   $d = 0.448$  \\
                  & $\chi^2_{\rm red}=1.3\times 10^{-4}$  &$\chi^2_{\rm red}=3.6\times 10^{-5}$  &$\chi^2_{\rm red}=2.4\times 10^{-4}$  &$\chi^2_{\rm red}=2.3\times 10^{-4}$  \\
\hline
Pd--Pd  & $a = -0.346$                                                 & $c = 1.40 \times 10^2$                                 & $a = -0.931$                                                &  $c = 17.9$  \\
                 &  $b = 1.08$                                                   & $d = 2.54$                                                      & $b = 3.36$                                                    &   $d = 1.13$  \\
                  & $\chi^2_{\rm red}=3.7\times 10^{-5}$  &$\chi^2_{\rm red}=2.6\times 10^{-5}$  &$\chi^2_{\rm red}=4.8\times 10^{-4}$  &$\chi^2_{\rm red}=6.0\times 10^{-4}$  \\
\hline
\end{tabular}
\end{center}
\label{tab:fit}
\end{table}%


\newpage

\end{document}